\documentclass[aps,prl,showpacs,twocolumn]{revtex4}
\usepackage{epsfig}
\usepackage{amsfonts}
\begin{document}


\title{Influence of detector motion in entanglement measurements with photons}


\author{Andr\'e G. S. Landulfo, George E. A. Matsas and Adriano C. Torres}
\affiliation{Instituto de F\'\i sica Te\'orica, Universidade Estadual Paulista,
         Rua Dr. Bento Teobaldo Ferraz, 271 - Bl. II, 01140-070, S\~ao Paulo, SP,
         Brazil}

\date{\today}


\begin{abstract}
We investigate how the polarization correlations of entangled photons described 
by wave packets are modified when measured by moving detectors. For this purpose, 
we analyze the Clauser-Horne-Shimony-Holt Bell inequality as a function of the 
apparatus velocity. Our analysis is motivated by future experiments with entangled 
photons designed to use satellites. This is a first step towards the 
implementation of quantum information protocols in a global scale. 
\end{abstract}


\maketitle

Entanglement plays a central role in quantum theory being one of its most 
distinguishing features~\cite{emaranhamento1, emaranhamento2}. It allows for 
the proof that no theory of local hidden variables 
can ever reproduce all of the predictions of quantum mechanics~\cite{Bell1}. 
As for applications, entanglement is crucial to quantum 
cryptography~\cite{cripto1, cripto2, cripto3}, teleportation~\cite{tele1}, 
dense coding~\cite{densecoding1} and to the conception of 
quantum computers (see, e.g., Refs.~\cite{comp0, comp2} and references therein). 
Currently, there is much interest in testing quantum mechanics for large space 
distances and eventually in implementing quantum information protocols in global 
scales~\cite{freespace0a, freespace0b, freespace1, 
freespace2, freespace3}. Photons seem to be  the ideal physical objects 
for this purpose. Since present technology limits the use of fiber optics 
in this context up to about 100~km~\cite{fiber}, the most viable alternative 
to go beyond happens to be free-space transmission using satellites and ground 
stations~\cite{satellite1, satellite4, satellite5}.
Here, rather than discussing the paramount technical challenges related to these 
experiments, we focus on an intrinsic physical restriction posed by the 
motion of the satellites when special relativity is taken into consideration. 
We address this issue by investigating the Clauser-Horne-Shimony-Holt~(CHSH) 
Bell inequality~\cite{CHSH69} for two entangled photons when one of the
detectors is boosted with some velocity. Hereafter we assume 
$\hbar=c=1$ unless stated otherwise.

Let us assume a system composed of two photons, $A$ and $B$, as emitted 
in opposite directions along the $z$ axis in a SPS cascade~\cite{peres95}. 
The polarization of photons $A$ and $B$ is measured along arbitrary 
directions as defined by the unit vectors $\hat{\bf a}_i$ and $\hat{\bf b}_j$ 
($i,j=1,2$), respectively, which are orthogonal to the $z$ axis. The distance 
between the two detectors is large enough to make both measurements causally 
disconnected. It is well known that the CHSH Bell inequality
\begin{equation}
| E (\hat{\bf a}_2, \hat{\bf b}_1) + 
  E (\hat{\bf a}_2, \hat{\bf b}_2) + 
  E (\hat{\bf a}_1, \hat{\bf b}_1) -
  E (\hat{\bf a}_1, \hat{\bf b}_2) | 
\leq 2 
\label{bellinequality}
\end{equation}
is satisfied for local hidden variable theories. Here 
\begin{equation} 
E (\hat{\bf a}_i, \hat{\bf b}_j) \equiv 
\lim_{N \to \infty} \frac{1}{N} \sum_{n=1}^N 
P^A(\hat{\bf a}_i) P^B (\hat{\bf b}_j)
\label{correlation}
\end{equation}
is the polarization correlation function obtained after an arbitrarily 
large number $N$ of experiments is performed, and $P^A(\hat{\bf a}_i) $ assume 
$+1$ or $-1$ values depending on whether the polarization of photon $A$ is
measured along $\hat{\bf a}_i$ or orthogonally to it, respectively, and analogously 
for $ P^B (\hat{\bf b}_j)$.

Now, we investigate inequality~(\ref{bellinequality}) in the context of quantum 
mechanics when we allow one of the detectors to move along the $z$ axis (say, 
carried by a satellite). Let us write the normalized state of a two-photon system
as~\cite{PT03, LT05}
\begin{equation}
|\psi\rangle = \sum_{s_A, s_B} 
               \int d^3 {k}_A d^3 {k}_B \,
               \psi_{s_A s_B} (\mathbf{k}_A, \mathbf{k}_B)
               |{\bf k}_A, {\bf \hat \epsilon}^{s_A}_{{\bf k}_A} \rangle 
               |{\bf k}_B, {\bf \hat \epsilon}^{s_B}_{{\bf k}_B} \rangle 
\label{est}
\end{equation}
where 
$$
\langle {\bf k'}_X|{\bf k}_X \rangle  
= 
\delta(\mathbf{k'}_X - \mathbf{k}_X), 
\;\;\;\;
\langle  {\bf \hat \epsilon}^{s_X}_{{\bf k}_X} 
| {\bf \hat \epsilon}^{s'_X}_{{\bf k}_X} \rangle 
= \delta_{s_X, s'_X}, 
$$
and
\begin{equation}
\sum_{s_A, s_B} \int d^3 {k}_A d^3 {k}_B  
| \psi_{s_A s_B} (\mathbf{k}_A, \mathbf{k}_B) |^2 = 1.
\label{norm3}
\end{equation}
Here, $X=A, B$ distinguishes between both particles, 
$k_X = ( \left \| \mathbf{k}_X \right \| , \mathbf{k}_X )$
are the corresponding four-momenta,
and $s_X=\pm 1$ labels two orthogonal helicity 
eigenstates $|{\bf \hat \epsilon}^{s_X}_{{\bf k}_X}\rangle$
for fixed three-momentum ${\bf k}_X$. We note that 
$|{\bf \hat \epsilon}^{s_X}_{{\bf k}_X}\rangle$ is associated
with the complex three-vector
\begin{equation}
{\bf \hat \epsilon}^{s_X}_{{\bf k}_X}
=
R(\hat{\bf k}_X){\bf \hat \epsilon}^{s_X}_{z},
\label{epsilon+-}
\end{equation}
where
$
{\bf \hat \epsilon}^{s_X}_{z} \equiv (1/\sqrt{2})(1, i s_X , 0)
$ 
are orthonormal vectors in the $x \bot y$ plane and 
$R(\hat{\bf k}_X)$ is the matrix which rotates   
$\hat{\bf z}=(0, 0, 1)$ into 
$$
\hat{{\bf k}}_X
={\bf k}_X/\|{\bf k}_X \| 
\equiv (\sin \theta \cos \phi, \sin \theta \sin \phi, \cos \theta)
$$
with $\theta, \phi$ being the usual spherical angles. 
Next, by using $|{\bf \hat \epsilon}^{s_X}_{{\bf k}_X}\rangle$, we
define a new pair of  normalized 
states~\cite{PT03}
\begin{equation}
|{\hat {\bf e}}_x({\bf k}_{X}) \rangle 
=\frac{x_{+}({\bf k}_X)| {\bf \hat \epsilon}^{+}_{{\bf k}_X}\rangle 
+x_{-}({\bf k}_X)| {\bf \hat \epsilon}^{-}_{{\bf k}_X}\rangle}{[|x_{+}({\bf k}_X) |^2 
+ |x_{-}({\bf k}_X) |^2]^{1/2}}, 
\label{ex}
\end{equation}
and
\begin{equation}
|{\hat {\bf e}}_y({{\bf k}_X}) \rangle
=\frac{y_{+}({\bf k}_X)| {\bf \hat \epsilon}^{+}_{{\bf k}_X}\rangle
+y_{-}({\bf k}_X)| {\bf \hat \epsilon}^{-}_{{\bf k}_X}\rangle}{
[|y_{+}({\bf k}_X) |^2 + |y_{-}({\bf k}_X) |^2]^{1/2}},
\label{ey}
\end{equation}
associated with the unit three-vectors $\hat{\bf e}_x({\bf k}_{X})$ and 
$\hat{\bf e}_y({\bf k}_{X})$ which  (i)~are the closest ones to 
$\hat{\bf x}=(1,0,0)$ and $\hat{\bf y}=(0,1,0)$, respectively, and 
(ii)~are contained in the plane orthogonal to ${\bf k}_X $. 
Here
\begin{eqnarray}
&& x_{\pm}({\bf k}_X) 
\equiv \frac{1}{\sqrt{2}}(\cos \theta \cos \phi \pm i \sin \phi), 
\label{x_pm}\\
&& y_{\pm}({\bf k}_X) 
\equiv \frac{1}{\sqrt{2}}(\cos \theta \sin \phi \mp i \cos \phi),
\label{y_pm}
\end{eqnarray}
and we note that 
$\hat{\bf e}_x({\bf k}_{X})$ and $\hat{\bf e}_y({\bf k}_{X})$ 
do not have to be mutually orthogonal.
By using Eqs.~(\ref{ex})-(\ref{ey}), the horizontal and 
vertical polarization states can be defined as
\begin{equation}
|H_X\rangle = 
\int d^3 k_X f_{{\bf p}_X}({\bf k}_X)   |{\bf k}_X, {\hat {\bf e}}_x({{\bf k}_X})\rangle 
\end{equation}
and
\begin{equation}
|V_X\rangle= 
\int d^3 k_X f_{{\bf p}_X}({\bf k}_X)   |{\bf k}_X, {\hat {\bf e}}_y({{\bf k}_X}) \rangle, 
\end{equation}
respectively, where the function $f_{{\bf p}_X}({\bf k}_X)$ gives the photon momentum
dispersion. By imposing that the dispersion is restricted to the $x\bot y$ plane 
and described by a Gaussian function, we write 
\begin{equation}
|f_{\mathbf{p}_X}(\mathbf{k}_X)|^2 = 
\pi^{-1} w^{-2} \delta(k^{z}_X - p_X^z) 
e^{-(k^r_{X}/w)^2} \;\; (w > 0),
\end{equation}
where 
$k^r_{X} \equiv \sqrt{(k^x_{X})^2 + (k^y_{X})^2\,}$ and
we assume that $\mathbf{p}_A = - \mathbf{p}_B = ( 0 , 0, |p|)$ 
since photons $A$ and $B$ move in opposite directions along 
the $z$ axis. 

Let us now assume that our two-photon entangled system is prepared in 
the state
\begin{equation}
|\psi \rangle 
= 
\frac{1}{\sqrt{2}}(|H_A\rangle \otimes |H_B\rangle 
+ |V_A\rangle\otimes |V_B\rangle)
\label{sps}
\end{equation}
and investigate the polarization correlations when the detector that measures, say, 
photon~$A$ is carried by a satellite with three-velocity ${\bf v}=(0, 0, v)$, 
while the other one, which measures photon~$B$, lies at rest at the ground station. 
This is important to note that each detector will see the state $|\psi \rangle$ 
in their proper frames unitarily transformed as~\cite{H68, W96}
\begin{equation}
|\psi \rangle \to |\psi'\rangle = 
U_A(\Lambda)\otimes I_B |\psi \rangle,
\label{psi'}
\end{equation}
where $I_B$ is the identity operator which acts in the Hilbert space associated 
with particle $B$ and
\begin{eqnarray}
&& U_A (\Lambda )  |{\bf k}_A, {\bf \hat \epsilon}^{s_A}_{{\bf k}_A} \rangle 
=
[ (\Lambda \ k_A)^0/k_A^0 ]^{{1}/{2}}  
\nonumber \\
&& \times \sum_{s'_A= \pm 1} D_{s'_A s^{}_A} (\Lambda, {\bf k}_A) 
 |\Lambda {\bf k}_A, {\bf \hat \epsilon}^{s'^{}_A}_{\Lambda {\bf k}_A} \rangle.
\label{U}
\end{eqnarray}
Here 
\begin{equation}
D_{s'^{}_A s_A }(\Lambda, {\bf k}_A)=
\exp{[-is'_A \Theta(\Lambda, {\bf k}_{A})]}\delta_{s'_A s^{}_A}
\label{wigner}
\end{equation}
is the Wigner rotation, 
where $\Theta(\Lambda, {\bf k}_{A})$ is a phase factor~\cite{GBA03,LPT03}. 
We note that ${\Lambda {\bf k}_A}$ denotes the
spatial part of the four-vector ${\Lambda k_A}$. For our particular choice
where the satellite moves along the $z$ direction with velocity $v$, the 
corresponding boost matrix $\Lambda$  is
$$
\Lambda_{\cal Z}= 
\left( \begin{array}{cccc} 
\cosh \alpha & 0 & 0 & \sinh \alpha\\
0 & 1 & 0&0 \\
0 & 0 & 1 & 0\\  
\sinh \alpha & 0& 0 & \cosh \alpha \\
\end{array} \right)
$$ 
with
$
\alpha \equiv - \tanh^{-1} v,
$
in which case 
$\Theta(\Lambda_{\cal Z}, {\bf k}_A)=0$.
By using Eqs.~(\ref{sps}), (\ref{psi'}) and~(\ref{U}), we obtain
\begin{equation}
|\psi' \rangle = \frac{1}{\sqrt{2}}(|H'_A\rangle \otimes |H_B\rangle + |V'_A\rangle\otimes |V_B\rangle),
\label{sps'}
\end{equation}
where
\begin{eqnarray}
&&
|H'_A\rangle= \int d^3 k_A \sqrt{{([{\Lambda_{\cal Z}}}]^{-1} k_A)^0/k_A^0}\; 
f_{{\bf p}_A} ([{{\Lambda_{\cal Z}}}]^{ -1}{\bf k}_A)  \nonumber \\
&&\times \frac{x_{+}([{{\Lambda}_{\cal Z}}]^{ -1}{\bf k}_A)|{\bf k}_A, {\bf \hat \epsilon}^{+}_{{\bf k}_A}\rangle+x_{-}([{\Lambda_{\cal Z}]^{-1}}{\bf k}_A)|{\bf k}_{A}, {\bf \hat \epsilon}^{-}_{{\bf k}_A}\rangle}{[|x_{+}([{{\Lambda_{\cal Z}}}]^{-1}{\bf k}_A) |^2 
+ |x_{-}([{{\Lambda}_{\cal Z}}]^{-1}{\bf k}_A) |^2]^{1/2}}, 
\nonumber \\
&& \\
&&|V'_A\rangle= \int d^3 k_A \sqrt{{([{\Lambda}_{\cal Z}}]^{-1} k_A)^0/k_A^0}\; 
f_{{\bf p}_A} ([{{\Lambda}_{\cal Z}}]^{ -1}{\bf k}_A)  \nonumber \\
&&\times \frac{y_{+}([{{\Lambda}_{\cal Z}}]^{ -1}{\bf k}_A)|{\bf k}_A, {\bf \hat \epsilon}^{+}_{{\bf k}_A}\rangle+y_{-}([{\Lambda_{\cal Z}}]^{-1}{\bf k}_A)|{\bf k}_{A}, {\bf \hat \epsilon}^{-}_{{\bf k}_A}\rangle}{[|y_{+}([{\Lambda_{\cal Z}]^{-1}}{\bf k}_A) |^2 
+ |y_{-}([{{\Lambda}_{\cal Z}}]^{-1}{\bf k}_A) |^2]^{1/2}}.
\nonumber \\
\end{eqnarray}

Next, we restrict the photon polarization measurements to the $x \bot y$ plane. 
This is convenient, hence, to define the operators 
\begin{eqnarray}
\! P^{X}_{x x} = 
| {\hat{x}_X} \rangle \langle {\hat{x}_X}| \otimes I^X_{ \bf{k} }, 
&& 
P^{X}_{x y} = 
| {\hat{x}_X} \rangle \langle {\hat{y}_X}| \otimes I^X_{ \bf{k} }, 
\label{P1}
\\
\! P^{X}_{y y} = 
| {\hat{y}_X} \rangle \langle {\hat{y}_X}| \otimes I^X_{ \bf{k} } ,
&& 
 P^{X}_{y x} = 
| {\hat{y}_X} \rangle \langle {\hat{x}_X}| \otimes I^X_{ \bf{k} }, 
\label{P2}
\end{eqnarray}
where  $I^X_{\bf{k}}$ is the identity operator acting in the momentum 
space of particle $X$ and
\begin{equation}
| \hat{x}_X \rangle
\equiv x_{+} ({\bf k}_X)|{\bf \hat \epsilon}^{+}_{{\bf k}_X}\rangle 
+ x_{-} ({\bf k}_X)|{\bf \hat \epsilon}^{-}_{{\bf k}_X}\rangle 
+ x_{l} ({\bf k}_X)|{\bf \hat \epsilon}^{l}_{{\bf k}_X}\rangle
\end{equation}
and
\begin{equation}
| \hat{y}_X \rangle
\equiv y_{+} ({\bf k}_X)|{\bf \hat \epsilon}^{+}_{{\bf k}_X}\rangle 
+ y_{-} ({\bf k}_X)|{\bf \hat \epsilon}^{-}_{{\bf k}_X}\rangle 
+ y_{l} ({\bf k}_X)|{\bf \hat \epsilon}^{l}_{{\bf k}_X}\rangle
\end{equation}
are associated with the unit vectors $\hat{\bf x}=(1, 0, 0)$ 
and $\hat{\bf y}=(0, 1, 0)$, respectively.
We recall that 
$ x_\pm ({\bf k}_X )$ 
and 
$y_\pm ({\bf k}_X )$
are given in Eqs.~(\ref{x_pm}) and~(\ref{y_pm}), respectively,
and 
$x_{l}({\bf k}_X) \equiv \hat{\bf x} \cdot \hat{{\bf k}}_X$,
$y_{l}({\bf k}_X) \equiv \hat{\bf y} \cdot \hat{{\bf k}}_X$.
In order to span a complete basis, we have introduced an unphysical 
longitudinal polarization state
$|{\bf \hat \epsilon}^{l}_{{\bf k}_X}\rangle$
associated with the three-vector 
${\bf \hat \epsilon}^{l}_{{\bf k}_X} \equiv \hat{{\bf k}}_X$,
as in Ref.~\cite{PT03}. 
Now, we use Eqs.~(\ref{P1})-(\ref{P2}) to introduce the operator
\begin{equation}
\sigma^{X}_{\varphi}
=
(P^{X}_{xx}-P^{X}_{yy})\cos{2 \varphi} 
+ (P^{X}_{xy}+P^{X}_{yx})\sin{2 \varphi},
\end{equation}
which will be useful further to compute the left-hand side of the CHSH 
Bell inequality~(\ref{bellinequality}). The eigenvalues $+ 1$ and $-1$ 
of the operator $\sigma^{X}_{\varphi}$ correspond to polarization 
eigenstates associated with directions tilted by angles $\varphi$ 
and $\varphi + \pi/2$ with respect to the $x$ axis, respectively. 
The correlation between the polarization measurements for the two particles 
$A$ and $B$ associated with directions defined by the angles $\varphi$ and 
$\varpi$, respectively, is given by
\begin{equation}
\langle \sigma^{A}_{\varphi}\otimes \sigma^{B}_{\varpi} \rangle_{\Psi}
=\langle \Psi | \sigma^{A}_{\varphi}\otimes \sigma^{B}_{\varpi}| \Psi \rangle,
\end{equation} 
where $|\Psi\rangle$ is the state of the two-photon system.
\begin{figure}[t]
\begin{center}
\includegraphics[height=0.25\textheight]{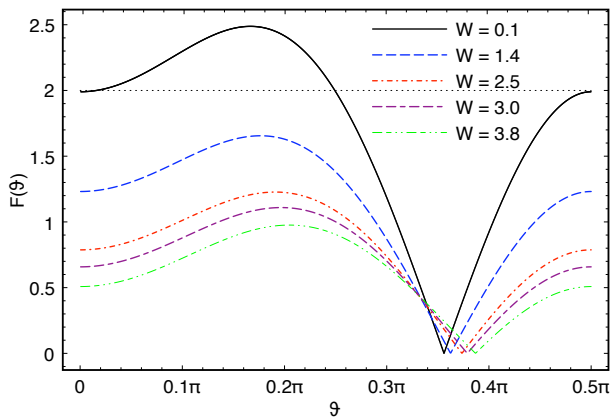}
\caption{(Color online) $F(\vartheta)$ is plotted as a
function of $\vartheta$ assuming $\alpha \to \infty$ for 
different values of the wave packet width properly normalized:
$W=w/|p|$. The larger the wave packets the smaller the
polarization correlations.}
\label{largura}
\end{center}
\end{figure}

For our purposes, this is enough to consider the case where 
$ \hat{\bf a}_2 = \hat{\bf b}_1 = \hat{\bf x}$.
By assuming that the unit vectors $\hat{\bf a}_1$ and 
$\hat{\bf b}_2 $ are counter-clockwisely and clockwisely rotated by an
angle $\vartheta$ with respect to the $x$ axis, respectively,
the left-hand side of Eq.~(\ref{bellinequality}) 
\begin{equation}
F(\vartheta) \equiv
| E (\hat{\bf a}_2, \hat{\bf b}_1) + 
  E (\hat{\bf a}_2, \hat{\bf b}_2) + 
  E (\hat{\bf a}_1, \hat{\bf b}_1) -
  E (\hat{\bf a}_1, \hat{\bf b}_2) | 
\end{equation}
is cast in the form
\begin{equation}
F(\vartheta) = 
|\langle 
\sigma^{A}_{0}\otimes \sigma^{B}_{0} 
+\sigma^{A}_{0}\otimes \sigma^{B}_{-\vartheta} 
+\sigma^{A}_{\vartheta}\otimes \sigma^{B}_{0}
-\sigma^{A}_{\vartheta}\otimes \sigma^{B}_{-\vartheta} 
\rangle_{\psi'}|,
\label{final}
\end{equation}
where the two-photon state $| \psi' \rangle$ is given in Eq.~(\ref{sps'}). 

\begin{figure}[t]
\begin{center}
\includegraphics[height=0.25\textheight]{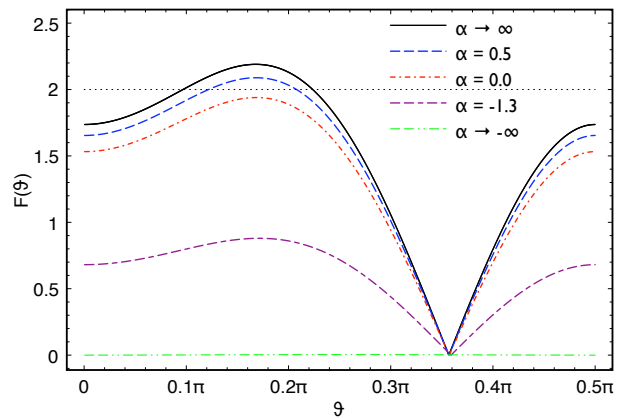}
\caption{(Color online) $F(\vartheta)$  is plotted as a
function of $\vartheta$ with $W= 0.6$ for different 
values of the apparatus velocity. Note that $F(\vartheta)$
drops as $\alpha$ decreases.}
\label{alfa}
\end{center}
\end{figure}

\begin{figure}[t]
\begin{center}
\includegraphics[height=0.25\textheight]{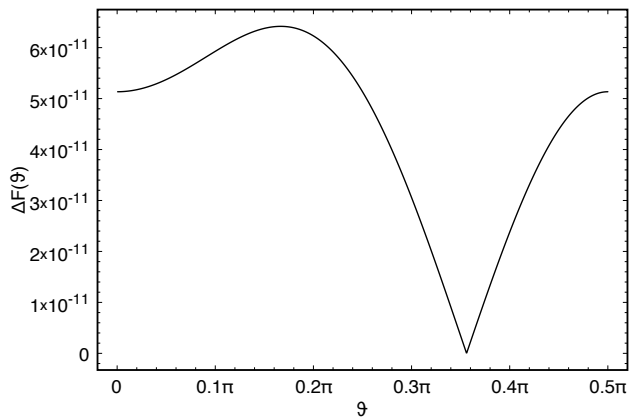}
\caption{(Color online) The graph exhibits 
$\Delta F(\vartheta) = F(\vartheta)-F_0(\vartheta)$. 
$F_0(\vartheta)$ is  obtained by imposing that both detectors 
lie at rest, while realistic values are used to calculate 
$F(\vartheta)$: we take the mean velocity of  the International 
Space Station to fix $\alpha = 2.6\, \times 10^{-5}$ and present 
technology for the production of entangled photons to fix 
$W  = 10^{-3}$.}
\label{diferenca}
\end{center}
\end{figure}

Next, we perform a numerical investigation of Eq.~(\ref{final}). 
As a consistency check, we have firstly verified that the standard 
CHSH Bell inequality, where $F(\vartheta)|_{\rm max} = 2.5$, is 
recovered for $\alpha =0$ and $w =0$. 
We recall that $\alpha< 0 $ and $\alpha > 0$ correspond to the cases 
where photon $A$ and the corresponding detector move towards the same 
and opposite directions, respectively.  In Fig.~\ref{largura} we exhibit 
how $F(\vartheta)$ is sensitive to the width of the photon wave packet 
properly normalized: $W \equiv w/|p|$. The plot assumes 
$\alpha \to \infty$ but the same pattern is verified for any other
fixed $\alpha$ (including $\alpha =0$). We see that the 
larger the wave packet the more $| \psi \rangle$ gets mixed
in polarization once momentum degrees of freedom are ignored and,
thus, the smaller the polarization correlation. 
In Fig.~\ref{alfa}, we plot $F(\vartheta)$ 
for different velocities of the moving detector assuming $W=0.6$. 
For large enough $|\alpha|$ ($\alpha<0$), we have that $F(\vartheta)$ 
is arbitrarily small in the whole domain. This shows how important the 
detector motion can be to polarization measurements when the velocity is high 
enough~\cite{PST02,LM09}. In order to understand the pattern observed in 
Fig.~\ref{alfa}, we note that as $\alpha$ decreases, photon $A$ 
becomes more redshifted according to the moving detector. As a consequence, 
$W$, which quantifies the wave dispersion normalized by the 
photon energy, ``looks like" larger in the detection frame. 
Hence, from Fig.~\ref{largura}, $F(\vartheta )$ should indeed drop 
as $\alpha$ decreases. In Fig.~\ref{diferenca}, we plot 
$\Delta F(\vartheta) = F(\vartheta)-F_0(\vartheta)$,
where $F_0(\vartheta)$ is  obtained by imposing that both detectors lie at rest
while realistic values are used to calculate $F(\vartheta)$: 
we take the mean velocity of  the International Space Station,
$v \approx 7.7 \, \times 10^{3} {\rm m/s}$, to fix $\alpha = 2.6\, \times 10^{-5}$  
and present technology for the production of entangled photons to fix 
$W = 10^{-3}$~\cite{width}.

Theoretical studies on the influence of the detector velocity 
in entanglement measurements is demanded by new perspectives of using 
satellites in quantum information experiments. Some laboratory 
effort to verify the influence of the detector motion in Bell 
inequalities using photons can be found in the literature.
In Ref.~\cite{SGZS02}, Stefanov, Zbinden, Ginsin and Suarez used an 
energy-time entangled photon pair state finding no 
signal for the influence of the detector motion in their results.
Although we cannot make any positive statement about their results
because we assume a distinct entangled state here, this is 
quite fair to expect from Figs.~\ref{alfa} and~\ref{diferenca} 
that any signal of the detector velocity would only be obvious for very 
relativistic systems. Furthermore, Fig.~\ref{largura} shows that 
the influence of the detector motion may be quite damped by using 
sharp enough wave packets ($W \ll 1$). In particular, for $w=0$ 
the detector velocity has no influence at all in $F(\vartheta)$. 
Incidentally, this may be an useful information for future applications 
of quantum protocols in a global scale. Although for present technology, 
detector motion effects should not play a dominant role as suggested by
Fig.~\ref{diferenca}, this will not be probably the case in the 
future when more precision will be attained. This is 
worthwhile to recall that the Global Positioning System would not work 
if the tiny desynchronization between satellite and ground antennas 
were not corrected by General Relativity formulas~\cite{A03} derived 
80 years earlier. 

\begin{acknowledgments}
The authors are indebted to Dr. R. Serra for calling their attention on 
the new experimental trends using entangled photons and satellites. 
A. L. and A. T. acknowledge full support from Funda\c c\~ao de Amparo 
\`a Pesquisa do Estado de S\~ao Paulo and Coordena\c c\~ao de Aperfei\c coamento 
de Pessoal de N\'ivel Superior, respectively. G. M. acknowledges partial 
support from Conselho Nacional de Desenvolvimento Cien\-t\'\i fico e 
Tecnol\'ogico and Funda\c c\~ao de Amparo \`a Pesquisa do Estado de 
S\~ao Paulo.  
\end{acknowledgments}

\end{document}